\documentclass{emulateapj}

\newcommand{\bvrino}{\protect\hbox{$BV\!RI$}}

\newcommand{\bvri}{\protect\hbox{$BV\!RI$} }

\newcommand{\jhk}{\protect\hbox{$JHK$}}
\newcommand{\about}{$\sim\!\!$~}

\newcommand{\be}{\begin{displaymath}}
\newcommand{\ee}{\end{displaymath}}

\def\lsim{\hbox{\rlap{\raise 0.425ex\hbox{$<$}}\lower 0.65ex\hbox{$\sim$}}}
\def\gsim{\hbox{\rlap{\raise 0.425ex\hbox{$>$}}\lower 0.65ex\hbox{$\sim$}}}

\newcommand{\kms}{km~s$^{-1}$}

\newcommand{\mkms}{{\rm \; km\;s^{-1}}}

\shorttitle{Optical Observations of SN~2006jc}
\shortauthors{Foley et al.}

\begin{document}

\title{SN~2006jc:  A Wolf-Rayet Star Exploding in a Dense He-Rich
Circumstellar Medium}

\author{Ryan J. Foley\altaffilmark{1}, Nathan Smith\altaffilmark{1},
Mohan Ganeshalingam\altaffilmark{1}, Weidong Li\altaffilmark{1}, Ryan
Chornock\altaffilmark{1}, and Alexei V. Filippenko\altaffilmark{1}}

\altaffiltext{1}{Department of Astronomy, University of California,
Berkeley, CA 94720-3411; {rfoley, nathans, mganesh, weidong, chornock,
alex}@astro.berkeley.edu}

\begin{abstract} 
We present optical photometry and spectra of the peculiar Type Ib
supernova (SN) 2006jc.  Strong and relatively narrow \ion{He}{1}
emission lines indicate the progenitor star exploded inside a dense
circumstellar medium (CSM) rich in He.  An exceptionally blue
apparent continuum persists from our first spectrum obtained 15 days after
discovery through our last spectrum \about1 month later.  Based on the
presence of isolated \ion{Fe}{2} emission lines, we interpret the blue
``continuum'' as blended, perhaps fluorescent, Fe emission. One or two
of the reddest \ion{He}{1} line profiles in our spectra are double 
peaked, suggesting that the CSM has an
aspherical geometry.  The \ion{He}{1} lines that are superposed on the
blue continuum show P-Cygni profiles, while the redder \ion{He}{1}
lines do not, implying that the blue continuum also originates from an
asymmetric mass distribution.  The He-rich CSM, aspherical geometry,
and line velocities indicate that the progenitor star was a WNE
Wolf-Rayet (WR) star.  A recent (2 years before the SN), coincident,
luminous outburst similar to those seen in luminous blue variables
(LBVs) is the leading candidate for the dense CSM.  Such an eruption
associated with a WR star has not been seen before, indicating that
the progenitor star may have recently transitioned from the LBV phase.
We also present unpublished spectral and photometric data on
SN~2002ao which, along with SN~1999cq, is very similar to SN~2006jc.
We propose that these three objects may represent a new and distinct class
of SNe arising from WR progenitors surrounded by a dense CSM.
\end{abstract}



\section{Introduction}\label{s:intro}

The most massive stars end their lives as core-collapse supernovae
(SNe).  The broad category of core-collapse SNe is divided
spectroscopically based on the presence or absence of H and He, with
the sequence from strong H, to He and weak H, to only He, to 
lacking both H and He being (respectively) Types II, IIb, Ib, and Ic
\citep[see][for a review]{Filippenko97}.
These correspond to progenitor stars with progressively decreasing H
and He envelopes, with the mass loss caused by either stellar winds or
mass transfer to a companion star.  Type IIn SNe \citep{Schlegel90}
are objects with narrow H emission lines, which are the result of the
SN ejecta interacting with a dense circumstellar medium (CSM)
\citep{Chugai94}.

Wolf-Rayet (WR) stars are evolved, core He-burning stars near
the end of their lives.  They are thought to be the descendants of
stars that begin their lives with initial masses $>$$40 M_{\sun}$,
which have already shed their H envelopes during a luminous blue
variable (LBV) stage \citep{Smith06}, exposing their He cores
\citep[for a review, see][]{Abbott87}.  During this stage in stellar
evolution, the mass loss from stellar winds is \about$10^{-5}
M_{\sun}$ yr$^{-1}$, about 100 times higher than that of other O and
B stars \citep{Crowther06}.  Although the WR stage lasts only \about$5
\times 10^{5}$ years \citep{Abbott87}, a star can lose several solar
masses of material during this stage.  These facts suggest that WR
stars are possible progenitors of stripped-envelope (Types IIb through
Ic) SNe.

\citet{Matheson00:99cq} showed that SN~1999cq had intermediate-width
(FWHM $\approx 2000 \mkms$) He emission lines similar to the
hydrogen lines of SNe~IIn, but
lacking H lines.  They suggested that SN~1999cq was interacting with a
CSM of dense He, but little or no H.  Until now, only one other
supernova, SN~2002ao \citep{Martin02}, has been identified as having
intermediate-width He emission lines
\citep{Filippenko02}.

SN~2006jc was discovered in \object{UGC 4904} by amateur astronomers
\citep{Nakano06} on 2006 Oct.\ 9.75 (UT dates are used throughout
this paper).  A non-detection was reported on Sep.\ 22, suggesting
that the SN was discovered shortly after explosion.  Soon after
detection, several groups obtained spectra of SN~2006jc and noted the
presence of \ion{He}{1} emission lines, but did not associate it with
SN~1999cq \citep{Crotts06, Fesen06:06jc1}.  \citet{Fesen06:06jc2}
later reclassified SN~2006jc as Type Ia, before \citet{Benetti06}
noted its similarity to SNe~1999cq and 2002ao.

In addition to optical photometry and spectroscopy presented in this
Letter, UV and X-ray data were obtained with the {\it Swift}
satellite \citep{Brown06}, with a positive X-ray detection, probably
indicating CS interaction.  Further X-ray observations with {\it
Chandra} \citep{Immler06} confirm the {\it Swift} X-ray results.

Detailed study of the well-observed SN~2006jc presents an opportunity
to examine the progenitor of a rare but important core-collapse SN
spanning the gap between SNe IIn and Ib, sampling the environment of a
WR star at the end of its stellar life, and helping to clarify what
appears to be a new subclass of SNe.  Here we present optical
photometry and spectroscopy and discuss the implications of those
data.  We also present observations of SN~2002ao in the interest of
completeness, but leave a detailed analysis of all the data for a
future paper.


\section{Observations}\label{s:obs}

Once the unusual nature of SN~2006jc was noted, we began a monitoring
campaign that consisted of \bvri photometry with the Katzman Automatic
Imaging Telescope \citep[KAIT;][]{Filippenko01} and optical
spectroscopy with the Kast spectrograph \citep{Miller93} mounted on
the Lick Observatory 3-m Shane telescope, the LRIS spectrograph
\citep{Oke95} mounted on the Keck I 10-m telescope, and the DEIMOS
spectrograph \citep{Faber03} mounted on the Keck II 10-m telescope.
The multi-band photometry began on Oct.\ 11 while our first spectrum
was obtained on Oct.\ 24. The photometry in this paper continues
through Dec.\ 16 while our last spectrum in this paper was obtained on
Nov.\ 24.

All spectral data were reduced using standard techniques
\citep[e.g.,][]{Foley03}.
Using our own IDL routines, we fit spectrophotometric
standard star spectra to flux-calibrate our data and remove telluric
lines \citep{Wade88, Matheson00:93j}.
Photometric data were obtained with KAIT and the 1-m Nickel telescope
at Lick Observatory.  Magnitudes were measured in the Kron-Johnson
\bvri system using the point-spread function (PSF)-fitting photometry
software \citep{Stetson87} in the IRAF\footnote{IRAF: the Image
Reduction and Analysis Facility is distributed by the National Optical
Astronomy Observatories, which is operated by the Association of
Universities for Research in Astronomy, Inc. (AURA) under cooperative
agreement with the National Science Foundation (NSF).} DAOPHOT
package, as multi-color template images required for galaxy
subtraction are not available and the SN is still visible in KAIT
data.  Since SN~2006jc is bright and reasonably isolated from its
faint host galaxy, galaxy subtraction is not necessary, and simple
PSF-fitting provides us a reasonable approach to reduce the photometry
at this early time.  The instrumental magnitudes for SN~2006jc derived
this way are calibrated with several local standard stars based on a
calibration from the photometric night of Oct.\ 21 with the Nickel
telescope.


\section{Results}\label{s:results}

\begin{figure}
\epsscale{1.20}
\rotatebox{90}{
\plotone{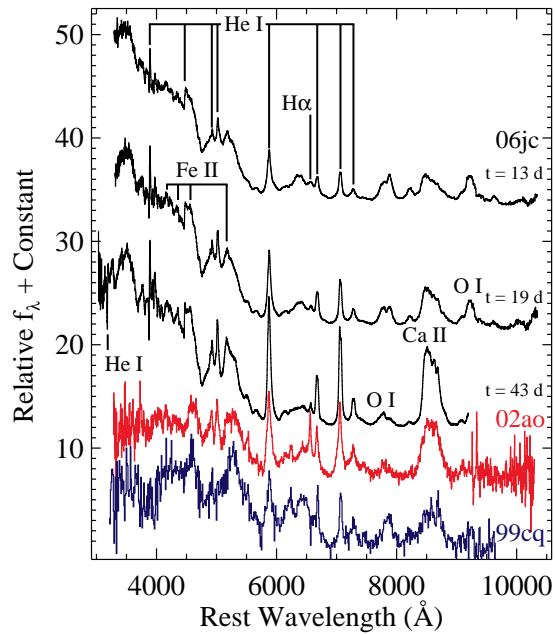}}
\caption{Spectra of SN~2006jc with some line identifications.  The
black curves are SN~2006jc (from top to bottom, the spectra were
obtained on 2006 Oct.\ 24.5, 2006 Oct.\ 30.6, and 2006 Nov.\ 23.6;
dates shown are relative to our first photometric observations), the
red curve is SN~2002ao obtained on 2002 Feb.\ 21.5, corresponding to
$t = 27$ d (all dates relative to our first photometric observation,
which, as seen in Figure~\ref{f:lc}, should be close to maximum
brightness) with the Lick 3-m telescope, and the blue curve is
SN~1999cq \citep{Matheson00:99cq} at $t = 20$ d.  A recession velocity
of 1670 \kms \citep{Nordgren97}, 1539 \kms \citep{Koribalski04}, and
8200 \kms \citep{Matheson00:99cq} has been removed from the spectra of
SNe~2006jc, 2002ao, and 1999cq, respectively.  The \ion{He}{1},
\ion{Fe}{2}, and \ion{Ca}{2} emission lines of SN~2006jc increase
relative to the continuum with time.  The blue continuum increases
relative to the red continuum with time.  SN~2006jc is similar to
SNe~2002ao and 1999cq.  The main difference is the very blue continuum
of SN~2006jc as well as a few lines, most notably the line at \about
6355~\AA.  However, the continuum differences are likely the result of
reddening for SNe~2002ao and 1999cq, while the line differences may
reflect the observations being obtained at slightly different
epochs.}\label{f:spec}
\end{figure}

\begin{figure}
\epsscale{1.2}
\rotatebox{90}{
\plotone{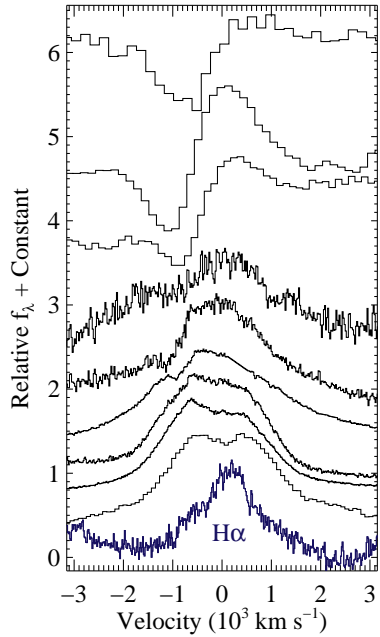}}
\caption{\ion{He}{1} lines of SN~2006jc from 2006 Nov.\ 23
(lower resolution) and 2006 Nov.\ 24 (higher resolution).  The lines
from top to bottom are \ion{He}{1} $\lambda 3188$, $\lambda 3889$,
$\lambda 4471$, $\lambda 4922$, $\lambda 5016$, $\lambda 5876$,
$\lambda 6678$, $\lambda 7065$,  $\lambda 7281$, and a line (in blue)
we identify as likely H$\alpha$.  The lines with $\lambda < 4900$ \AA\
exhibit P-Cygni profiles.  The \ion{He}{1} lines with $\lambda >
4900$ \AA\ have broad peaks, with \ion{He}{1} $\lambda 7281$ showing a
double-peaked profile.  The H$\alpha$ profile is much narrower than
that of the \ion{He}{1} lines.}\label{f:vel}
\end{figure}

We present our low-resolution spectra of SN~2006jc in
Figure~\ref{f:spec}.  In all spectra, we are able to identify
intermediate-width \ion{He}{1}, H$\alpha$, \ion{O}{1}, \ion{Ca}{2},
and \ion{Fe}{2} emission lines.  There are no photospheric P-Cygni
profiles typically found in early SN~Ib/c spectra \citep{Matheson01};
the spectra seem to consist of two continuum components (red and
blue) and intermediate-width ($1000 < v < 4000~\mkms$) emission lines.
The redder He lines show a intermediate-width emission component (FWHM
$\approx 3000~\mkms$) similar to, but slightly wider than, that of
SNe~IIn \citep{Filippenko97}.  As seen in Figure~\ref{f:vel}, the
bluer He lines show narrow P-Cygni profiles, with the absorption
minimum blueshifted by roughly $-1000~\mkms$.  The emission components
of the lines with P-Cygni profiles are narrower (FWHM $= 1000~\mkms$)
than the pure emission components, presumably 
because their blueshifted emission
is self-absorbed.  The only discernible characteristic between the
groups of lines with the two different types of line profiles is
wavelength; in particular, the groups are not distinguished by singlet
or triplet state.  The line intensity ratios of the \ion{He}{1} lines
evolve with time, which may be an indication of changing densities or
NLTE effects.

The spectra all show a relatively flat continuum redward of \about
5500 \AA, but a very steep blue continuum shortward of \about 5500
\AA, with a minimum $B-R$ color of $-0.45$ mag.  We have identified a
few isolated \ion{Fe}{2} lines in the spectra of SN~2006jc (see
Figure~\ref{f:spec}), suggesting that there are many other \ion{Fe}{2}
lines, mostly at short wavelengths.  It is likely that most of the
apparent blue ``continuum'' consists of 
blended Fe lines. \citet{Turatto93} suggested that
\ion{Fe}{2} emission lines create the excess blue continuum of
SN~1988Z. The blending of the Fe lines makes disentangling the lines
difficult, and detailed modeling is necessary to fully interpret our
data.  However, the amplitudes of the undulations in the blue
continuum become larger with time, despite no significant change in
the continuum shape.  If the apparent continuum is the result of
blended Fe lines, the increase of these amplitudes follows the
expected result of \ion{Fe}{2} lines increasing relative to the
continuum (as seen with isolated \ion{Fe}{2} lines).

There are still many features that we are unable to identify, most
notably that centered at 6357~\AA\, which may be \ion{Si}{2}
$\lambda6355$, with FWHM $\approx 6200~\mkms$.  This feature has some
substructure and is wider than other emission lines, and thus is
likely a blend of several lines.  While most emission lines increase
relative to the continuum with time, this feature decreases
dramatically relative to the continuum and disappears by our Nov.\ 23
spectrum, which is likely an excitation effect.  Since the \ion{Fe}{2}
lines increase relative to the continuum with time, it is most likely
some other element.  The features at 7881~\AA\ (somewhat blended with
\ion{O}{1} $\lambda7774$), 8215~\AA\, and 9360~\AA\ exhibit the same
behavior and width, and may be of the same species.

\begin{figure}
\epsscale{1.0}
\rotatebox{0}{
\plotone{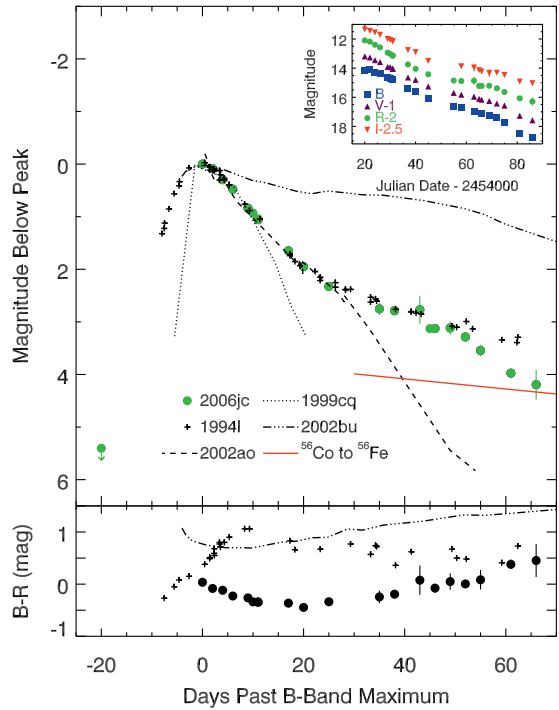}}
\caption{({\it top large panel}): $R$-band light curve of SN~2006jc
with comparison light curves of SNe~1994I (Ic), 1999cq, 2002ao, and
2002bu (IIn).  The light curve of SN~2002bu is very different from the
others, having a slower rise and decline, and a slight plateau.  All
other light curves appear similar for \about10 days past $R$ maximum,
after which SN~1999cq declines much faster than the other three.  At
\about25 days, SNe~2006jc and 1994I slow their decline, while
SN~2002ao continues its decline.  At \about55 days, SN~2006jc begins a
faster decline, separating from SN~1994I.  \citet{Arkharov06} note a
brightening in \jhk from days \about30 to 55, corresponding to the
plateau in the $R$ band.  The expected light curve from $^{56}$Co
decay, corresponding to 0.0098 mag day$^{-1}$, is plotted as the solid
red line.  ({\it inset}): \bvri light curves of SN~2006jc.  The SN
declines rapidly with $\Delta m_{15} (B) = 1.14$. The early (late)
decline rates are 0.088 (0.100), 0.087 (0.092), 0.099 (0.066), and
0.092 (0.054) mag day$^{-1}$ for \bvrino, respectively. ({\it bottom
panel}):  The $B-R$ color curve of SN~2006jc with comparison curves of
SNe~1994I (Ic) and 2002bu (IIn).  SN~2006jc is very blue with $B-R =
-0.45$ mag on JD 2454040.}
\label{f:lc}
\end{figure}

The slope of the blue continuum of SN~2006jc does not change much
with time.  The emission components of all He lines increase relative
to the continuum with time, while some other species, including
\ion{O}{1}, decrease with time.  The \ion{Ca}{2} IR triplet also
increases with time, consistent with other SNe.  The evolution of the
blue continuum is apparent in Figure~\ref{f:lc}.  The $B$-band light
curve declines more slowly than the other bands, which is unusual for
a SN light curve.  This supports the argument that the blue continuum
is the result of a process not typically seen in SNe.

SN~2006jc has maximum absolute magnitudes of $-17.8$, $-17.7$,
$-17.8$, and $-18.1$ for \bvrino, respectively, making it
as luminous as SN~1994I \citep{Richmond96}.  Assuming that nearly all
of its luminosity comes from $^{56}$Ni to $^{56}$Co to $^{56}$Fe
decay, the fact that SNe~1994I and 2006jc have similar luminosities
suggests they created similar $^{56}$Ni masses, \about0.07 $M_{\sun}$
\citep{Nomoto94}.

The early-time decline rates of SNe~2006jc, 2002ao, and 1999cq are
much faster than those of normal SNe~IIn (see Figure~\ref{f:lc} for a
comparison to SN~IIn 2002bu).  The early light curves of
SNe~IIn are powered mainly by circumstellar interaction and their slow
declines at early times indicate a massive CSM.  Despite showing
obvious features of circumstellar interaction in their spectra,
SNe~2006jc, 2002ao, and 1999cq all decline very quickly, suggesting an
extremely dense, but low-mass and clumpy CSM. The late-time decline rates 
of these objects are also very fast, indicating either a small $^{56}$Ni
mass or a small ejecta mass.


\section{Discussion}\label{s:discussion}

\subsection{Wolf-Rayet Progenitor in a Dense, He-Rich CSM}\label{ss:he}

The strength of the intermediate-width \ion{He}{1} emission lines in
SN~2006jc relative to other SNe~Ib is indicative of a dense CSM rich
in He.  Similar to the H lines seen in SN~IIn spectra, the
He lines of SN~2006jc are the result of the SN
ejecta interacting with a dense CSM.  The \ion{He}{1} lines seen only
in emission have FWHM $\approx$ 2000--3000~\kms, and the
absorption minimum of the P-Cygni profiles is \about 1000~\kms,
velocities typical of WR winds \citep{Abbott87}.  The identification
of H$\alpha$ suggests that the progenitor recently ejected material
containing at least some H into its surroundings, perhaps meaning that
the progenitor recently evolved from a LBV to a WNE WR star
\citep{Abbott87}.  Additionally, the smaller width of the H$\alpha$
line is consistent with a different ejection mechanism from that of
the He.  LBVs typically have slower expansion speeds than WR stars,
indicating that the H might have been ejected during the LBV stage,
while the He was ejected during the WR stage.  \citet{Matheson00:99cq}
did not see any H$\alpha$ in the spectrum of SN~1999cq, but the
low signal-to-noise ratio of the spectrum makes the measurement
ambiguous. The spectrum of SN~2002ao does show relatively strong
H$\alpha$ (see Figure~\ref{f:spec}), suggesting that its CSM contains
more H than that of SN~2006jc.

As noted by \citet{Nakano06}, there was a bright ($M = -14.0$ mag)
outburst coincident with the position of SN~2006jc in Oct.\ 2004,
which was initially thought to be an LBV outburst similar to SN~1961V
\citep{Goodrich89}.  Unless it was a mere coincidence, it seems that
the progenitor of SN~2006jc suffered an event analogous to the
non-terminal eruptions of LBVs shortly before final core collapse,
ejecting a He-rich shell with which the SN ejecta are now
interacting.  The short time between these events may have 
far-reaching implications for the late evolution of massive stars,
beyond what we can discuss here; however, it is worth noting that
\citet{Chugai04} determined that SN~IIn~1994W had an outburst
\about1.5 years prior to explosion.  Equally surprising is that this
type of event may have happened in a star inferred to be a WN star;
while such outbursts are known to occur in LBVs that still have their
H envelopes, no such variability has been documented in WN stars.  It
may hint that the progenitor of SN~2006jc had just recently
transitioned from the LBV to the WN stage, supporting the notion that
LBV mass loss facilitates the onset of the WR phase.  This would seem
consistent with the presence of H in its CSM.

\subsection{Blue Continuum}\label{ss:bluecont}

The slow $B$-band decline relative to the $R$ band, the steep blue
continuum, and the P-Cygni profiles for \ion{He}{1} all suggest that
the emission at wavelengths shorter than \about5500 \AA\ arises from a
mechanism that is different from that which produces the red
continuum.  We identified some isolated \ion{Fe}{2} lines, suggesting
that many other \ion{Fe}{2} lines are present in the spectrum, but are
blended and appear as a pseudo-continuum because of their large line
widths.  Since we do not see other high-excitation lines, this
emission might be provided by fluorescence and not collisional
excitation.

\citet{Matheson00:99cq} noted that SN~1999cq had a bluer continuum
than normal SNe~Ic.  They also noted that SN~1999cq had $E(B-V)
\lesssim 0.45$ mag, and argued that $E(B-V) \lesssim 0.25$ mag.
Similarly, SN~2002ao may be extinguished.  Correcting for reasonably
small reddening, the continua of both SNe~1999cq and 2002ao have
similar colors to that of SN~2006jc, indicating that these events may
also have a strong blue continuum.  The main difference
between the spectra is the strong \ion{Ca}{2} H\&K absorption in
SN~1999cq.

\subsection{Aspherical Geometry}\label{ss:aspherical}

The double-peaked nature of some \ion{He}{1} lines suggest a complex,
asymmetric CSM.  Unlike SNe~IIn, the light curve of SN~2006jc declined
very quickly, suggesting that there is little mass in the CSM.
However, the strong emission lines indicate a very dense CSM.  These
facts are compatible if the CSM is asymmetric or clumpy.  Also, the
P-Cygni profiles seen only for the blue \ion{He}{1} lines also
indicate asymmetry, because the blue continuum is absorbed by the He
gas, whereas the red continuum (presumably from the inner SN debris)
is not.

A simple way to explain these observables is supernova ejecta
interacting with a bipolar CSM, which causes the double-peaked nature
of the He lines.  The He-rich CSM, which was likely the result of  the
recent (2004) outburst, may be asymmetric because of either rapid
rotation in the progenitor or strong interaction with a binary
companion.

\begin{acknowledgments}

Some of the data presented herein were obtained at the W. M. Keck
Observatory, which was made possible by the generous financial support
of the W. M. Keck Foundation.  We thank the Keck and Lick Observatory
staffs for their assistance.  This research was supported by NSF grant
AST-0607485 and the TABASGO Foundation.  KAIT was made possible by
generous donations from Sun Microsystems, Inc., the Hewlett-Packard
Company, AutoScope Corporation, Lick Observatory, the NSF, the
University of California, and the Sylvia \& Jim Katzman Foundation.
\end{acknowledgments}

{\it Facilities:} \facility{KAIT}, \facility{Nickel}, \facility{Shane
(Kast Double spectrograph)}, \facility{Keck:I (LRIS)},
\facility{Keck:II (DEIMOS)}

\clearpage

\end{document}